\documentclass[a4paper,11pt]{article}
\usepackage{Proceeding_template/pos}
\usepackage{float}
\usepackage{wrapfig}
\usepackage{multicol}

\title{Gamma-Ray Polarization Results of the POLAR Mission and Future Prospects}
 \ShortTitle{GRB Polarimetry with POLAR and POLAR-2}

\author*{Merlin Kole}
\affiliation{DPNC, University of Geneva, Switzerland}
\emailAdd{merlin.kole@unige.ch}

\forColl{POLAR\footnote{\url{https://www.astro.unige.ch/polar/collaboration}} \& POLAR-2\footnote{\url{https://www.unige.ch/dpnc/polar-2}}}

\abstract{Despite over 50 years of Gamma-Ray Burst (GRB) observations many open questions remain about their nature and the environments in which the emission takes place. Polarization measurements of the GRB prompt emission have long been theorized to be able to answer most of these questions. The POLAR detector was a dedicated GRB polarimeter developed by a Swiss, Chinese and Polish collaboration. The instrument was launched, together with the second Chinese Space Lab, the Tiangong-2, in September 2016 after which it took 6 months of scientific data. During this period POLAR detected 55 GRBs as well as several pulsars. From the analysis of the GRB polarization catalog we see that the prompt emission is lowly polarized or fully unpolarized. There is, however, the caveat that within single pulses there are strong hints of an evolving polarization angle which washes out the polarization degree in the time integrated analysis. Building on the success of the POLAR mission, the POLAR-2 instrument is currently under development. POLAR-2 is a Swiss, Chinese, Polish and German collaboration and was recently approved for launch in 2024. Thanks to its large sensitivity POLAR-2 will produce polarization measurements of at least 50 GRBs per year with a precision equal or higher than the best results published by POLAR. POLAR-2 thereby aims to make the prompt polarization a standard observable and produce catalogs of the gamma-ray polarization of GRBs. Here we will present an overview of the POLAR mission and all its scientific measurement results. Additionally, we will present an overview of the future POLAR-2 mission, and how it will answer some of the questions raised by the POLAR results.}

\FullConference{37$^{\rm{th}}$ International Cosmic Ray Conference (ICRC 2021)\\
		July 12th -- 23rd, 2021\\
		Online -- Berlin, Germany}

\graphicspath{{./Pictures/}}

\begin{document}
\maketitle

\section{Introduction}

Gamma-ray bursts (GRBs) are the brightest electromagnetic transient phenomena in the Universe. Over 5 decades after their discovery, they remain one of the most studied phenomena in astrophysics. Such studies performed on these transient events have spanned many wavelengths, from radio wavelengths to TeV energies, and have revealed much about their origin. The initial phase of a GRB, referred to as the prompt phase is mostly visible in $\gamma$-rays and has a bimodal duration 
distribution, where the short GRBs have a typical duration of $t_{\rm GRB}\sim10^{-1}\,$s and the long GRBs typically last for $t_{\rm GRB}\sim30$s \citep{Kouveliotou+93}. The prompt emission is followed by a longer lasting afterglow mostly visible at lower energies. Thanks to measurements of this afterglow, which allow for precise localization of GRBs, it was possible to associate long GRBs with star forming regions in galaxies as well as supernovae events, linking them to the death of massive stars \cite{Galama+98}. Short GRBs were long theorized to be the results of neutron star mergers, a theory which was proven by the joint detection of gravitational waves and the short GRB 170817A \cite{Abbott+17}.

The prompt emission we observe is thought to originate from 2 jets of highly relativistic material emitted during the death of a massive star, in case of a long GRB, or during a neutron star merger event in case of a short GRB. The afterglow is then produced as the material in the jets interacts with the interstellar material around the progenitor. Although the general picture of these transients are fairly clear, the exact radiation mechanism responsible for the prompt emission and the morphology and composition of the jets remain poorly understood. 

Whereas spectrometry measurements of the prompt emission have provided a deep insight into the uncertainties raised above, it appears that spectrometry measurements alone will not be able to fully resolve these. For this purpose $\gamma$-ray polarization measurements of the prompt emission have long been proposed as a solution.

Measurements of the polarization degree (PD) as well as the angle will allow to distinguish between, for example, photospheric emission models and synchrotron emission models. Whereas further detailed measurements of the distribution of the PD as well as the evolution of the PD and polarization angle (PA) can answer further questions on for example the magnetic configuration of the jets. As the type of questions which can be resolved which such measurements is vast as well as complex, we refer the reader to \cite{Toma+11} and \cite{Gill+21} for a comprehensive overview of the model resolving power of $\gamma$-ray polarimetry.

\section{GRB Polarization Measurements by POLAR}

Due to the wealth of information which can be gained through $\gamma$-ray polarization measurements, a large number of attempts has been made to perform these over the last 2 decades. To date the polarization of a total of 31 GRBs has been published. For several of these GRBs different groups have either studied the same data and in one case the polarization was studied using data from 2 different instruments. Despite this effort no strong conclusions can be drawn based on these results for several reasons. The first is the large statistical errors in all the measurements resulting from the relatively low sensitivity of $\gamma$-ray polarimeters which is typically an order of magnitude lower than that of a $\gamma$-ray spectrometer. A second issue with the majority of the existing measurements stems from systematic errors. For a detailed overview of the polarization measurement performed prior to 2016 the reader is referred to \citep{McConnell+16}. 

In order to produce the first reliable and precise measurements, the last decade has seen the rise of the first dedicated well calibrated $\gamma$-ray polarimeters. The first of these was GAP, which led to publications of polarization measurements of 3 bright GRBs, one indicating a relatively low PD \cite{GAP1} while the other 2 favoured a PD of $\geq50\%$ \cite{GAP2}. These first 3 measurements indicated that GRB polarimetry was possible, however, both the statistical errors and the small sample size still did not allow to draw any strong conclusions. 

The success of GAP was followed up with the launch of the POLAR detector \cite{Produit+18}. A dedicated GRB polarimeter developed by a Chinese, Swiss and Polish collaboration. This plastic scintillator based instrument was launched as part of the Tiangong-2 spacelab in 2016 and took data for approximately half a year. During this period a total of 55 GRBs were detected as well as several pulsars (see \cite{HCL} for details on the pulsar analysis). 

\subsection{Time Integrated POLAR results}

Out of the 55 GRBs detected by POLAR, 5 were selected for an initial polarization study. The selection criteria applied were based on the brightness of the GRB, its incoming angle with respect to the zenith pointing direction of POLAR and the GRB should have been observed by another instrument.

\begin{wrapfigure}{l}{0.5\textwidth}
\includegraphics[width=0.9\linewidth]{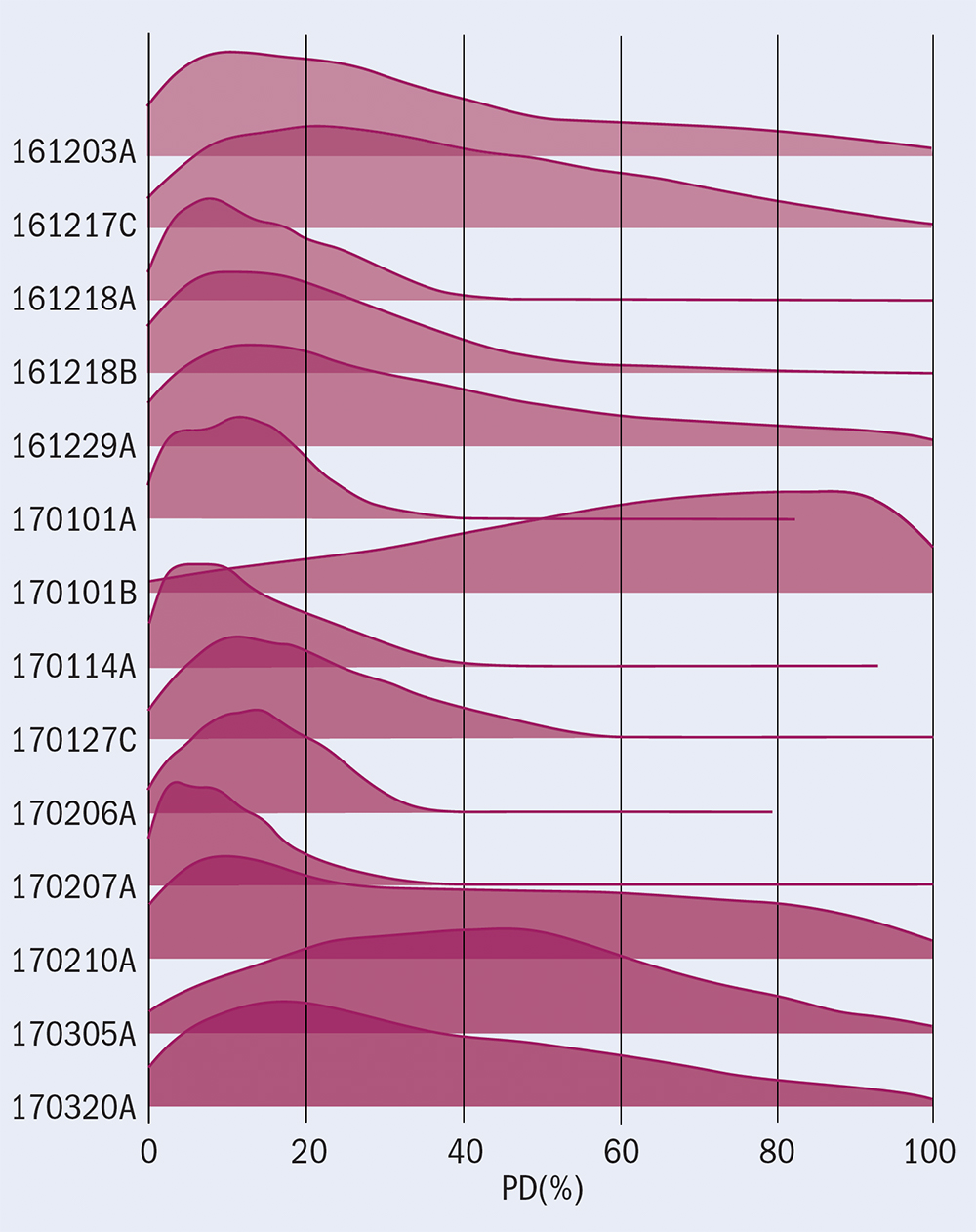}  \caption{Posterior distributions of the 14 GRBs analyzed in \cite{Kole+20}. The results appear to indicate a low or fully unpolarized emission.}
 \label{posterior_dist_POLAR}
\end{wrapfigure}

The initial criterion is placed to ensure a signal to background ratio to allow for statistically significant measurements. The second and third criteria were placed to reduce systematic uncertainties which are induced by uncertainties on the spectrum and location of the GRB, while the last criterion additionally ensures that $\gamma$-rays scattering off materials surrounding POLAR do not influence the final result. 

The 5 GRBs passing the criteria were all found to be consistent with a low PD and an unpolarized flux could not be excluded \cite{Zhang+19}. The analysis procedure was later improved in order to relax the GRB selection criteria. For this purpose an instrument independent analysis tool, described in \cite{Burgess+19}, was developed which apart from polarimetry data allows to include spectral data from other instruments, such as \textit{Fermi}-GBM and Swift-BAT. The spectrometry data  allowed to reduce the systematic errors in the polarization results stemming from uncertainties in the spectral parameters. Additionally, it allowed to study potential systematic errors in the POLAR response from, for example, scattering of photons from objects in the vicinity of POLAR. This study showed no significant systematic effects in the POLAR instrument response for any of the GRBs observed by both POLAR and \textit{Fermi}-GBM or Swift-BAT and indicated a good cross calibration of the three instruments \cite{Kole+20}. 

Thanks to this new analysis, GRBs only observed by POLAR could be included in the analysis as POLAR data could now be used to measure the spectrum directly. Additionally GRBs with larger incoming angles could be analyzed after verification that no additional systematics were present in the instrument response. Finally, thanks to further improvements in the analysis procedure the fluence limit was lowered, resulting in a total GRB sample of 14. The posterior distribution of the PD for all these GRBs is shown in figure \ref{posterior_dist_POLAR} \cite{Kole+20}. The data and instrument responses required for the analysis were made public in 2019\footnote{\url{https://www.astro.unige.ch/polar/grb-light-curves}}. Overall the results are fully compatible with an unpolarized or lowly polarized flux whereas polarization degrees above $\sim50\%$ are disfavoured.

During the period POLAR was active the AstroSat CZTI instrument also performed polarization measurements of GRBs \cite{Tanmoy+19}. Although not a dedicated polarimeter, the instrument was shown to be capable of performing such measurements on ground using on-axis polarizated beams \cite{AstoSat_calib}. The results presented by AstroSat for 11 GRBs in \cite{Tanmoy+19} all favour a polarization degree above $\sim50\%$ and are thereby in contrast with the POLAR results. Future joint analysis, using for example the tool presented in \cite{Burgess+19} are required to understand this discrepancy. Finally, the balloon-borne COSI experiment performed a polarization measurement of a GRB \cite{Lowell}. The analysis resulted in an upper limit of $46\%$.

\subsection{Time resolved results}

While most theoretical works on the polarization of the GRB prompt emission have focused on the time integrated polarization, predictions do exist on the evolution of the PD and PA during the GRB. Additionally, several measurements have been performed in the past (see \cite{McConnell+16} for an overview) which show indications of a potential evolution in both the PD and the PA during the burst.

The POLAR data was used to perform time resolved analysis on all 14 GRBs shown in figure \ref{posterior_dist_POLAR}. While for 12 of these GRBs no PD was found as well in the time resolved analysis, for the two Fast Rising Exponential Decay (FRED) like GRBs in the sample, hints of an evolution within the pulse were found \cite{Kole+20}. In both cases a PD of around $30\%$ was found with an evolving PA. A detailed analysis in \cite{Burgess+19} of one of these, 170114A, shows what appears to be a continuous evolution, however, a single $90^\circ$ change in PA cannot be excluded using the available statistics. Due to the limited statistics more detailed time resolved studies, as well as energy resolved studies, are not possible. 

\section{The POLAR-2 Instrument}

\begin{figure*}
\begin{multicols}{2}
\includegraphics[width=0.7\linewidth]{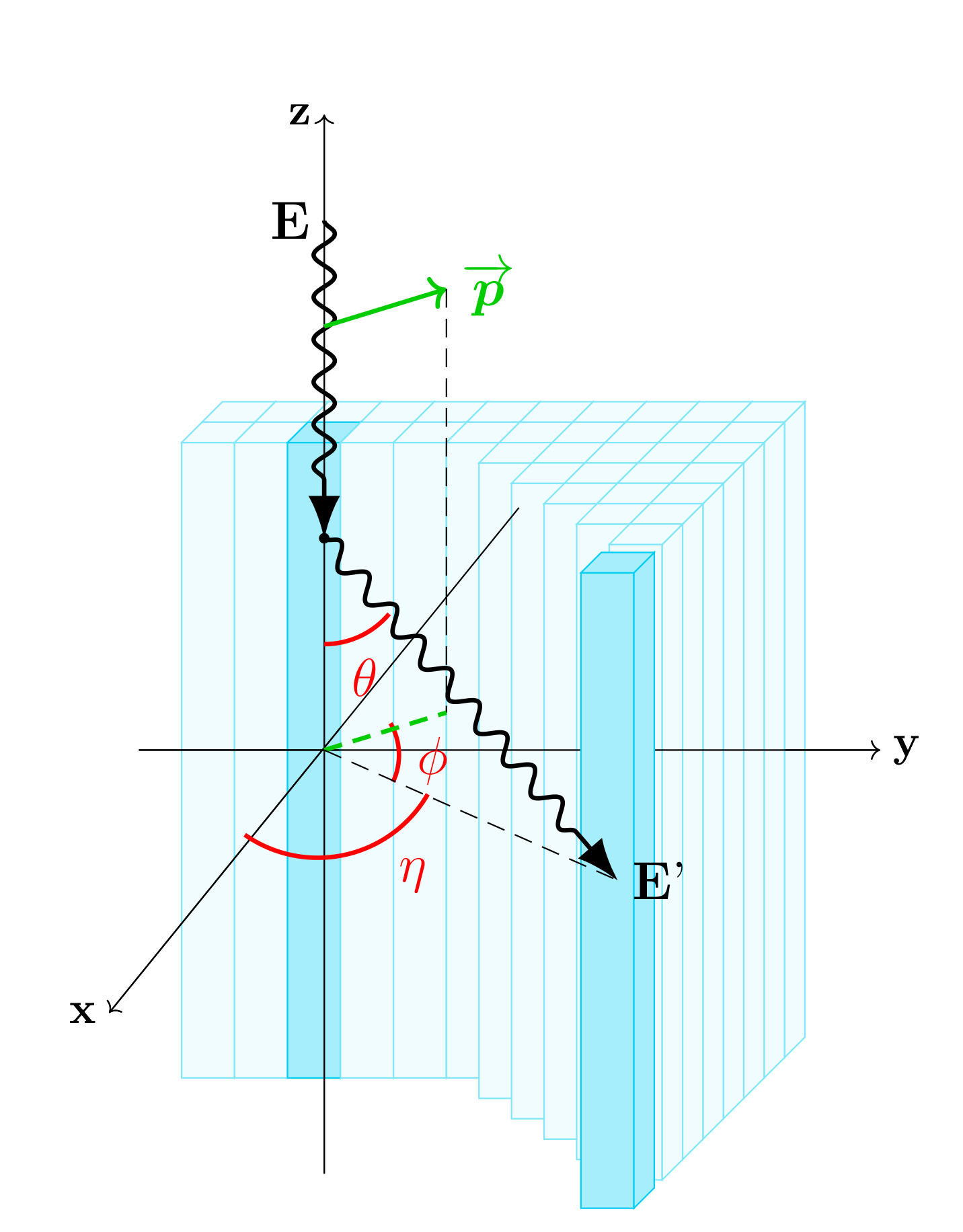}
 \caption{Illustration of the measurement principle of a polarimeter using Compton scattering. 
    The incoming $\gamma$-ray Compton scatters in one of the detector segments followed by a 
    photo-absorption (or second Compton scattering) interaction in a different segment. Using the 
    relative position of the two detector segments the Compton scattering angle can be calculated 
    from which, in turn, the polarization angle can be deduced.}
 \label{measurement_principle}
\includegraphics[width=0.6\linewidth]{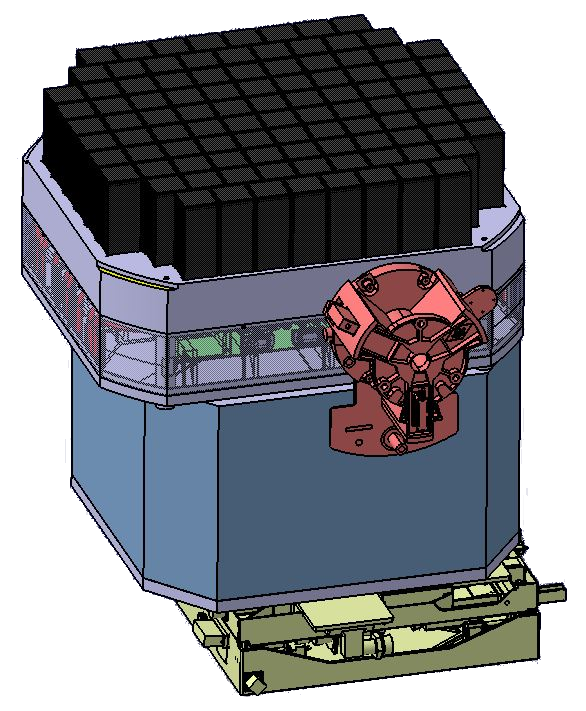}
 \caption{A schematic overview of the POLAR-2 design. The 100 detector modules (in black) are shown on top the aluminium grid which houses the front-end electronics of the modules. The aluminium grid is placed on top a larger structure which houses the back-end electronics, low voltage power supply and additionally serves as a radiator. In red a grabbing system for a robotic arm used during installation of the payload on the CSS is shown. The interface to the CSS, both mechanically and electrically is seen at the bottom in yellow. The height of the payload is $700\,\mathrm{mm}$ while the cross section is $590\times648\,\mathrm{mm^2}$. The total mass is foreseen to be 150 kg an the power consumption 300W.}
 \label{POLAR-2_design}
 \end{multicols}
\end{figure*}

The lack of conclusions which can be drawn from the time resolved as well as the time integrated studies show the need for a significantly more sensitive detector. For this purpose the POLAR-2 detector was proposed by Swiss (UniGe), Chinese (IHEP), Polish (NCBJ) and German (MPE) collaboration to be placed on the China Space Station (CSS). The instrument proposal was accepted in July 2019, resulting in a confirmed launch to take place in early 2024.

The principle used to measure polarization in POLAR-2 is equal to that used in POLAR. The azimuthal Compton scattering angle, $\phi$, which has a dependency on the polarization angle of the incoming $\gamma$-ray, is measured using a plastic scintillator array. This principle is illustrated in figure \ref{measurement_principle}. The first major improvement of POLAR-2 over POLAR regards the number of scintillators, which is increased from 1600 to a total of 6400, thereby increasing the effective area of the instrument by approximately a factor of 4. The 6400 bars are divided into groups of 64, each forming together with a SiPM array and its own front-end electronics (FEE) an individual detector module which is housed in a carbon fibre structure. The carbon fibre structure serves both for mechanical stability as well as a shielding of low energy charged particles.

The 100 detector modules are placed in an aluminium grid, as shown in figure \ref{POLAR-2_design}. The aluminium grid houses the FEE and the Peltier based cooling system of the SiPM arrays. The FEE of each module handles its trigger logic, the temperature regulated high voltage of the SiPM, the power of the Peltier element and sends trigger information and data to the back-end electronics. The back-end electronics (BEE) along with the low voltage power supply is placed in the large structure below the aluminium grid.  Apart from housing these components the structure is designed to have a large radiating surface allowing to have passive cooling of the instrument which is foreseen to consume 300 W. Additionally the structure holds a grabbing system (shown in red in figure \ref{POLAR-2_design}) which will be used by a robotic arm during the installation of POLAR-2 on the payload platform of the CSS. The BEE handles the powering of the full instrument, communication between the FEEs, combines the data from the FEEs into data packages and handles the communication with the CSS. The CSS has computing facilities allowing POLAR-2 data to be searched for transients and for quick alerts to be sent to ground based instruments based on these transient searches.

Several design improvements were made to increase the sensitivity of POLAR-2. Firstly the shape of the scintillator is optimized with respect to that used in POLAR. The scintillator bars are made shorter (from 176 mm to 125 mm) to improve the signal to background ratio and increase the sensitivity to polarization. Furthermore, the cross section of each bar increased from $5.8\times 5.8\,\mathrm{mm}^2$ to $5.9\times 5.9\,\mathrm{mm}^2$. A second major technological improvement is the change from 8x8 Multi Anode PMTs to arrays of SiPM which increases the light yield of each readout channel and reduces optical cross talk. This final change warrants an additional change in the scintillator shape to increase the scintillator readout surface which in POLAR was reduced due to optical cross talk. These changes result in an increase in an average light yield of $1.6\,\mathrm{photo-electrons/keV}$ compared to $0.3\,\mathrm{photo-electrons/keV}$ in POLAR. This results in a significant increase in the effective area at low energies. Overall, the effective area (for polarization events) is an order of magnitude larger compared to POLAR. Details on the design changes and their effects on the scientific performance are discussed in detail in \cite{NDA}. 

\subsection{Scientific Prospects}

\begin{figure}[H]
\centering
\includegraphics[height=0.37\textwidth]{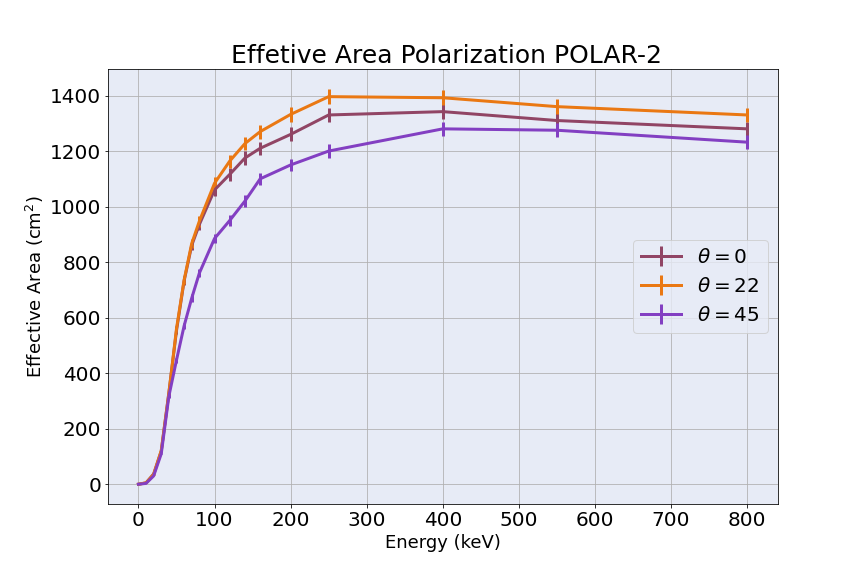}\includegraphics[height=0.37\textwidth]{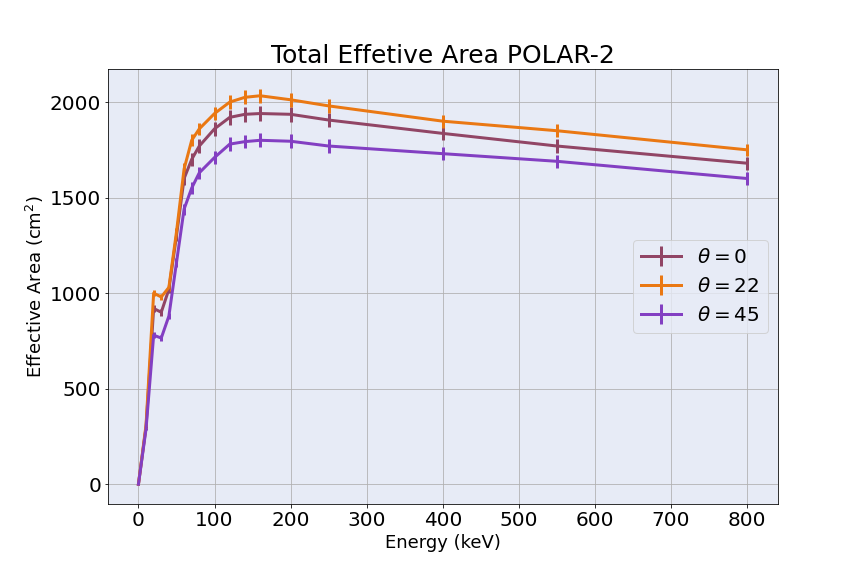}
 \caption{Effective area for polarization events as a function of energy for 3 different incoming angles on the left. On the right the effective area is shown for polarization + single bar trigger events.}
 \label{Effective_area}
\end{figure}

The scientific performance of POLAR-2 was simulated using the Geant4 framework \cite{Geant4}. The simulations are based on the performance found during the calibration  of the first detector modules assembled during 2021. Details of these measurements are provided in \cite{NDA}. The effective area was calculated for 2 different types of triggers. The first, the results of which are shown in the left in figure \ref{Effective_area} for different incoming angles $\theta$ of the photons, contains only triggers where there are energy depositions above the set lower energy threshold in at least 2 scintillator bars and have an energy deposition below the overflow of the low gain readout of the ASIC reading out the SiPM. The events which pass this criterion can be used for polarization studies. Apart from such events, POLAR-2 will also trigger when only a single energy deposition is recorded in the instrument. To limit dead time and reduce the daily data size, the threshold for such events is set higher at 7 photo-electrons, whereas for polarization events this is set at to around 4 photo-electrons. Although such single bar events cannot be used for polarization studies they can be used for spectral measurements, GRB localization and pulsar studies such as those described in \cite{HCL}. The effective area for all triggers, meaning both single bar and polarization events is shown on the right in figure \ref{Effective_area}.

\begin{figure}[H]
\centering
\includegraphics[height=0.55\textwidth]{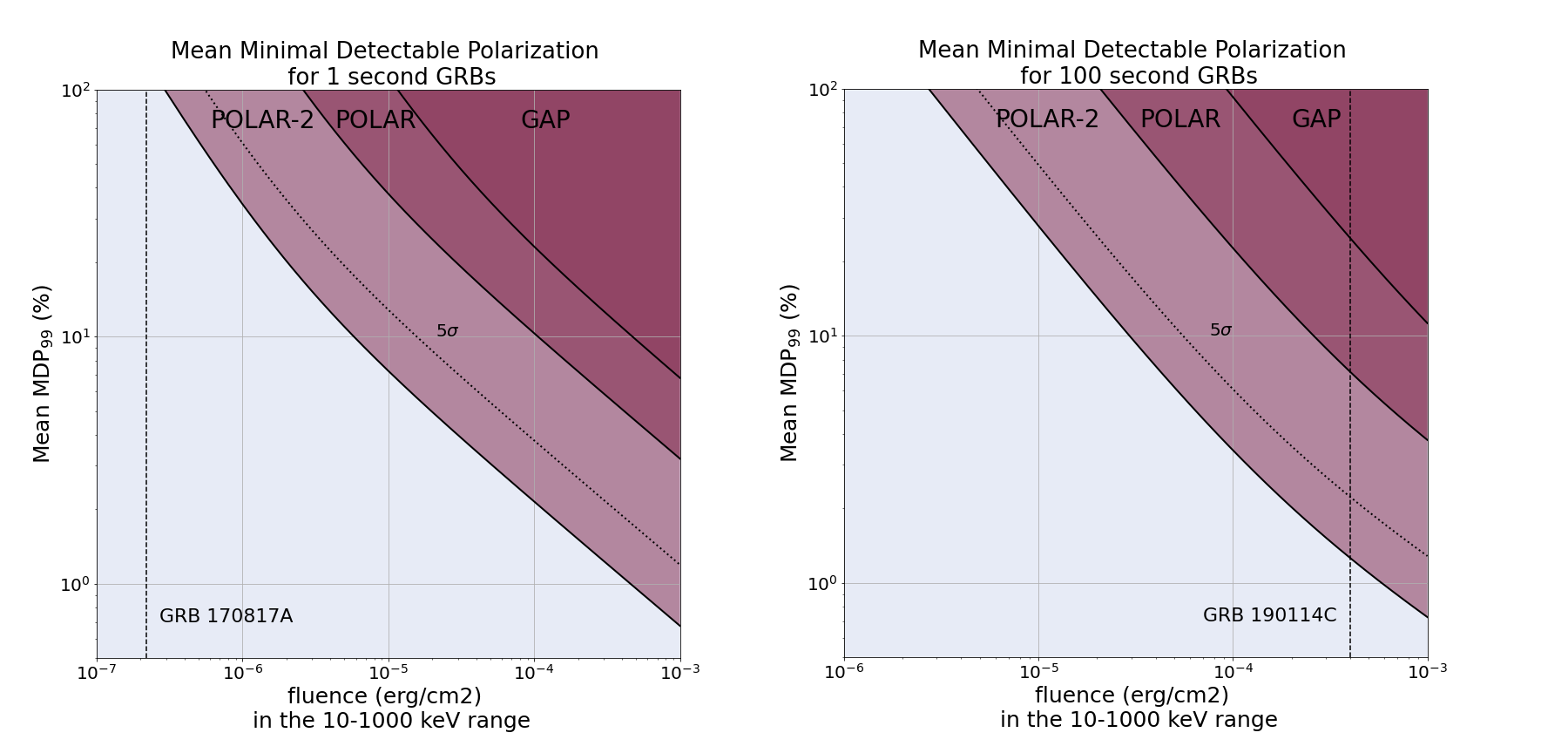}
 \caption{The mean minimal detectable polarization for $99\%$ confidence level (MDP averaged over PA) as a function of the fluence in the 10-1000 keV energy band for GAP, POLAR and POLAR-2 for both 1 second (left) and 100 second long (right) GRBs. The MDP for $5\sigma$ confidence is added using a dotted line for POLAR-2 as well. The fluences of 2 well known GRBs, the weak and short 170817A and the long and bright 190114C, are added as an indication.}
 \label{fluence}
\end{figure}

Based on these results, as well as the sensitivity to polarization, described in \cite{NDA}, the mean Minimal Detectable Polarization (MDP), meaning the MDP averaged over all polarization angles, for both typical long and short GRBs can be calculated as a function of the GRB fluence. The results are shown in figure \ref{fluence} along with the performance of POLAR and GAP. The figure indicates that POLAR was able to reject PD above $\sim50\%$ within half a year of data taking. POLAR-2 will not only be able to easily detect GRBs as faint as GRB 170817A, but will be able to perform polarization measurements of GRBs only slightly brighter than this one. Based on this performance, POLAR-2 will be able to probe the PD down to levels as low as $10\%$ for 10 GRBs per year and perform 50 measurements per year with a precision equal or better than the 5 most precise measurements by POLAR. More details on this performance can be found in an upcoming publication \cite{Gill+21}.

\section*{Acknowledgements}

We gratefully acknowledge the Swiss Space Office of the State Secretariat for Education, Research and Innovation (ESA PRODEX Programme) which supported the development and production of the POLAR-2 detector. M.K and N.D.A. acknowledge the support of the Swiss National Science Foundation. JMB acknowledges support from the Alexander von Humboldt Foundation. National Centre for Nuclear Research acknowledges support from Polish National Science Center under the grant UMO-2018/30/M/ST9/00757. We gratefully acknowledge the support from the National Natural Science Foundation of China (Grant No. 11961141013, 11503028), the Xie Jialin Foundation of the Institute of High Energy Physics, Chinese Academy of Sciences (Grant No. 2019IHEPZZBS111), the Joint Research Fund in Astronomy under the cooperative agreement between the National Natural Science Foundation of China and the Chinese Academy of Sciences (Grant No. U1631242), the National Basic Research Program (973 Program) of China (Grant No. 2014CB845800), the Strategic Priority Research Program of the Chinese Academy of Sciences (Grant No. XDB23040400), and the Youth Innovation Promotion Association of Chinese Academy of Sciences.

\clearpage
\section*{Full Authors List: POLAR-2 Collaboration}


\scriptsize
\noindent
M.~Kole$^1$,
N.~De~Angelis$^1$,
J.M.~Burgess$^4$,
F.~Cadoux$^1$,
J.~Greiner$^4$,
J.~Hulsman$^1$,
H.C.~Li$^2$,
S.~Mianowski$^3$,
A.~Pollo$^3$,
N.~Produit$^2$,
D.~Rybka$^3$,
J.~Stauffer$^1$,
J.C.~Sun$^5$,
B.B.~Wu$^5$,
X.~Wu$^1$,
A.~Zadrozny$^3$
and
S.N.~Zhang$^{5,6}$ \\

\noindent
$^1$University of Geneva, DPNC, 24 Quai Ernest-Ansermet, CH-1211 Geneva, Switzerland.\\
$^2$University of Geneva, Geneva Observatory, ISDC, 16, Chemin d’Ecogia, CH-1290 Versoix, Switzerland.\\
$^3$National Centre for Nuclear Research, ul. A. Soltana 7, 05-400 Otwock, Swierk, Poland.\\
$^4$Max-Planck-Institut fur extraterrestrische Physik, Giessenbachstrasse 1, D-85748 Garching, Germany.\\
$^5$Key Laboratory of Particle Astrophysics, Institute of High Energy Physics, Chinese Academy of Sciences, Beijing 100049, China.\\
$^6$University of Chinese Academy of Sciences, Beijing 100049, China.

\end{document}